\documentclass{article}
\usepackage{waspaa15,graphicx,url,times,mathrsfs}
\usepackage{amsmath,amsfonts,multirow,stmaryrd}
\usepackage{algorithm,algorithmic}
\usepackage[justification=centering]{caption}

\graphicspath{{figures/}}

\title{Phase reconstruction of spectrograms based on a model of repeated audio events}

\name{Paul Magron \qquad Roland Badeau \qquad Bertrand David \thanks{This work is partly supported by the French National Research Agency (ANR) as a part of the EDISON 3D project (ANR-13-CORD-0008-02) and the AIDA project (ANR-13-CORD-0001).}}
\address{Institut Mines-T\'{e}l\'{e}com, T\'{e}l\'{e}com ParisTech, CNRS LTCI, Paris, France \\ \texttt{<firstname>.<lastname>@telecom-paristech.fr}}

\begin{document}

\ninept
\maketitle

\begin{sloppy}

\begin{abstract}
Phase recovery of modified spectrograms is a major issue in audio signal processing applications, such as source separation. This paper introduces a novel technique for estimating the phases of components in complex mixtures within onset frames in the Time-Frequency (TF) domain. We propose to exploit the phase repetitions from one onset frame to another. We introduce a reference phase which characterizes a component independently of its activation times. The onset phases of a component are then modeled as the sum of this reference and an offset which is linearly dependent on the frequency. We derive a complex mixture model within onset frames and we provide two algorithms for the estimation of the model phase parameters. The model is estimated on experimental data and this technique is integrated into an audio source separation framework. The results demonstrate that this model is a promising tool for exploiting phase repetitions, and point out its potential for separating overlapping components in complex mixtures.
\end{abstract}


\section{Introduction}

A variety of audio signal processing techniques acts in the TF domain, exploiting the particular structure of music signals. For instance, the family of techniques based on Nonnegative Matrix Factorization (NMF) is often applied to spectrogram-like representations, and has proved to provide a successful and promising framework for audio source separation~\cite{Smaragdis2003}.

However, when it comes to resynthesizing time signals, obtaining the phase of the corresponding Short-Time Fourier Transform (STFT) is necessary. In order to produce perceptually satisfactory sounding signals, it is important to enforce \emph{consistency}, i.e. to obtain a complex-valued component that is close to the STFT of a time signal. In the source separation framework, a common practice consists in applying Wiener-like filtering (soft masking of the complex-valued STFT of the original mixture)~\cite{Fevotte2009}. However, this method does generally not lead to consistent components. Alternatively, a consistency-based approach is often used for phase recovery~\cite{Griffin1984}. That is, a complex-valued matrix is iteratively computed in order to maximize its consistency. A recent benchmark has been conducted to assess the potential of source separation methods with phase recovery in NMF~\cite{Magron2015}. It points out that consistency-based approaches provide poor results in terms of audio quality. Besides, Wiener filtering fails to provide good results when sources overlap in the TF domain. Thus, phase recovery of modified audio spectrograms is still an open issue~\cite{Gerkmann2015}.

Another approach to reconstruct the phase of a spectrogram is to use a phase model based on the analysis of mixtures of slowly-varying sinusoids~\cite{McAuley1986}. Contrary to consistency-based approaches using the redundancy of the STFT, this model exploits the natural relationships between adjacent TF bins. This approach is used in the phase vocoder algorithm~\cite{Flanagan1966}, where it is mainly dedicated to time stretching. In~\cite{Bronson2014}, a complex NMF framework with phase constraints based on sinusoidal modeling is introduced. In~\cite{Magron2015a}, the authors proposed to generalize this approach and have provided a phase unwrapping algorithm that has been applied to an audio signal restoration task. However, the knowledge of the phases within onset frames is required to initialize the unwrapping. Finally, relative phase offsets between partials have been exploited in a complex matrix decomposition framework in~\cite{Kirchhoff2014}.


In this paper, we propose to exploit the phase repetitions from one onset frame to another in order to estimate the phases of the complex-valued components composing a mixture. We model the phase of a given source within an onset frame as the sum of a reference phase and an offset which is linearly dependent on the frequency. We then derive a mixture model of complex components within onset frames. The phase parameters are estimated by means of two algorithms, relying on either a \emph{strict} or a \emph{relaxed} phase constraint.
Phase parameters estimation is performed on experimental signals. It is also combined to the linear unwrapping technique~\cite{Magron2015a} and integrated into an audio source separation framework. Contrary to consistency-based approaches, this technique is based on a sinusoidal model, thus it will be compared to the traditional Wiener filtering technique.
Note that the proposed reconstruction method can be used with any spectrogram factorization technique.

The paper is organized as follows. Section \ref{sec:model} presents the phase model. Section \ref{sec:model_est} is dedicated to the estimation of the phase parameters. Section \ref{sec:exp} describes several experiments that highlight the potential of this technique. Finally, section \ref{sec:conclu} draws some concluding remarks and prospects future directions.

\section{Repeating phase model}
\label{sec:model}

\subsection{Main concept}
Let us consider a time signal $x(n)$, $n \in \mathbb{Z}$. The expression of the STFT is, for each frequency channel $f \in \llbracket 0 ; F-1 \rrbracket$ and $t \in \mathbb{Z}$: 

\begin{equation}
X(f,t) = \sum_{n=0}^{N-1} x(n+tS) w(n) e^{-2i \pi \frac{f}{F} n },
\label{eq:stft}
\end{equation}
where $w$ is an $N$ sample-long analysis window and $S$ is the time shift (in samples) between successive frames.

We assume that $x$ represents a source that is activated \emph{twice}. The corresponding onset frame indexes in the TF domain are denoted by $t_1$ and $t_2$. The key idea here is that $X(f,t_1)$ and $X(f,t_2)$ are the Fourier transforms of two signals that are approximately equal up to a gain factor $\rho \in \mathbb{R}_+$ and a time delay $\eta \in \llbracket 0 ; N-1 \rrbracket $. In the TF domain, that leads to:

\begin{equation}
X(f,t_2) \approx  X(f,t_1) \rho e^{i \lambda f}, \text{ with } \lambda = \frac{2 \pi \eta}{F}.
\label{eq:red_mod}
\end{equation}

The family of NMF techniques exploits (\ref{eq:red_mod}) through its magnitude. Indeed, it models the spectrogram of a source as a spectral template that is activated over time with a variable gain factor. We propose here to exploit (\ref{eq:red_mod}) through its phase: the phase of a component within an onset frame $t_m$ is obtained by applying an offset to a \emph{reference phase} that only depends on the frequency:
 
\begin{equation}
\angle X(f,t_m) = \phi(f,t_m) \approx \psi(f) + \lambda (m) f,
\label{eq:phase_inv}
\end{equation}
where $\angle(.)$ denotes the complex argument. Since the reference phase is defined up to a delay, we can set for instance $\lambda(0) = 0$
to ensure that $\psi(f)$ is not ambiguously defined\footnote{However, when dealing with mixtures of sources in practical applications, observing an isolated source is no longer guaranteed, since audio sources often overlap in the TF domain. Then, both $\psi^k$ and $\lambda^k$ parameters must be estimated for any source $k$.}.

\subsection{Example}
We propose to investigate the validity of the model~(\ref{eq:phase_inv}) on a simple example. We consider signals made up of two occurrences of a piano note obtained from the Midi Aligned Piano Sounds (MAPS) database~\cite{Emiya2010a}. A gain factor is applied to the second occurrence. We then compute the phase difference between onset frames. According to the model, this difference is expected to be a linear function of the frequency. This procedure is illustrated in Figure \ref{fig:unwrap_error}.

\begin{figure}
\center
\includegraphics[scale=0.4]{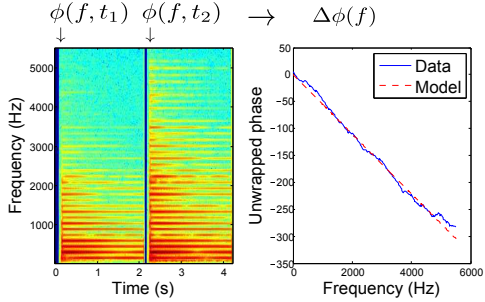}
\vspace{-0.5em}
\caption{Spectrogram with two onset frames (left) and phase difference between onset frames (right)}
\label{fig:unwrap_error}
\vspace{-1.5em}
\end{figure}

The relative error between the observed data and the modeled estimate is averaged over $30$ signals. A $1.5 \%$ error is obtained. Those first observations seem to assess the accuracy of our model, and justify to exploit phase repetitions by modeling the phase difference between onset frames as a linear function of the frequency.

\subsection{Mixture model}
Let us now consider the STFT $X \in \mathbb{C}^{F \times T}$ of a mixture of $K$ sources $X^k$, whose magnitudes and phases are denoted $A^k$ and $\phi^k$ respectively. We denote $t_m$ the $m$-th onset frame, $m \in \llbracket 0 ; M-1 \rrbracket$. We then define the onset matrix $Y \in \mathbb{C}^{F \times M} $:

\begin{equation}
Y(f,m) = X(f,t_m) = \sum_{k=1}^K  A^k(f,t_m) e^{ i \phi^k(f,t_m)}.
\label{eq:mod_mix}
\end{equation}

Incorporating the model (\ref{eq:phase_inv}) in (\ref{eq:mod_mix}) leads to the following mixture model: $\forall (f,m) \in \llbracket0,F-1\rrbracket \times \llbracket0,M-1\rrbracket$,

\begin{equation}
\hat{Y}(f,m) = \sum_{k=1}^K  A^k(f,t_m) e^{ i \psi^k(f)}  e^{i \lambda^k(m) f  }.
\label{eq:mod_inv}
\end{equation}

It is worth noting that this model reduces the dimensionality of the data within onset frames (assuming that $M>1$): the onset phases of the sources $\phi^k$ are represented by $KFM$ parameters, while our model uses only $K(F+M)$ parameters.

\section{Phase parameters estimation}
\label{sec:model_est}
We propose in this section a technique for estimating the phase parameters $\psi^k(f)$ and $\lambda^k(m)$ in model~(\ref{eq:mod_inv}). Magnitude parameters $A^k$ are assumed to be known. The estimation is performed through the minimization of a cost function. A first method consists in choosing the cost function as the squared Euclidean distance between the data and the model. We qualify this technique as \emph{strict} because the parameters are directly learned from the data.

However, when the data are no longer properly modeled by~(\ref{eq:mod_inv}), a strict constraint may be too restrictive to estimate the parameters. We then propose a method based on alternating the estimation of the phases of the components (\ref{eq:mod_mix}) and the estimation of the phase parameters (\ref{eq:mod_inv}). Drawing on previous work such as~\cite{Rigaud2013}, this leads to a method that we will qualify as \emph{relaxed}.

\subsection{Strict phase constraint}
\label{sec:est_strict}

In this section, we consider the following cost function:

\begin{equation}
\mathcal{C}_s = \sum_{f,m} | Y(f,m) - \sum_{k=1}^K  A^k(f,t_m) e^{ i \psi^k(f)} e^{i \lambda^k(m) f } |^2.
\label{eq:cost_strict}
\end{equation}

\noindent The parameter estimation is then performed in two steps.

\textbf{Estimation of \boldmath$\displaystyle \psi^k(f)$\unboldmath}. We calculate the partial derivative of $\mathcal{C}_s$ with respect to $\psi^k(f)$ and we seek $\psi^k(f)$ such that this derivative is zero. This leads to the following estimation:

\begin{equation}
\psi^k(f) = \angle \left( \sum_{m} B^k(f,m) A^k(f,t_m) e^{- i \lambda^k(m) f} \right),
\label{eq:strict_psi}
\end{equation}
where $B^k(f,m) = Y(f,m) - \sum_{l \neq k}  A^l(f,t_m) e^{ i \psi^l(f)} e^{i \lambda^l(m) f }$.

\textbf{Estimation of \boldmath$\displaystyle \lambda^k(m)$\unboldmath}. The problem of minimizing $\mathcal{C}_s$ with respect to $\lambda^k(m)$ becomes that of minimizing:

\begin{equation}
\tilde{\mathcal{C}}_s(k,m) = \sum_{f} | B^k(f,m) e^{- i \psi^k(f)} -  A^k(f,t_m) e^{i \lambda^k(m) f } |^2.
\label{eq:cost_strict_lambda_reform}
\end{equation}

Let us note $\beta^k(f,m) = B^k(f,m) e^{- i \psi^k(f)}$ and:
\begin{align*}
\underline{\beta}^k(m) &= [\beta^k(0,m),...,\beta^k(F-1,m)]^T,\\
\underline{\Lambda}^k(m) &= [1,e^{i \lambda^k(m)},...,e^{i \lambda^k(m)(F-1)}]^T,\\
 \underline{\alpha}^k(m) &= [A^k(0,t_m),...,A^k(F-1,t_m)]^T.
\end{align*}

\noindent Then, (\ref{eq:cost_strict_lambda_reform}) can be rewritten as:
\begin{equation}
\tilde{\mathcal{C}}_s(k,m) = || \underline{\alpha}^k(m) \odot \underline{\Lambda}^k(m) - \underline{\beta}^k (m) ||^2,
\label{eq:cost_strict_lambda_vec}
\end{equation}
where $||.||$ denotes the Euclidean norm and $\odot$ the Hadamard product. This problem can be solved by means of an adaptation of the ESPRIT algorithm~\cite{Hua2004}. Indeed, we observe that when $\tilde{\mathcal{C}}_s$ is zero, then (we partially omit the indexes $k$ and $m$ for more clarity):

\begin{equation}
\underline{\beta}_\downarrow^H \underline{\beta}_\uparrow
= (\underline{\alpha} \odot \underline{\Lambda})_\downarrow^H (\underline{\alpha} \odot \underline{\Lambda})_\uparrow
= \underline{\alpha}_\downarrow^H \underline{\alpha}_\uparrow e^{i \lambda^k(m)},
\end{equation}
where $\underline{v}_\downarrow$ (resp. $\underline{v}_\uparrow$) denotes the vector obtained by removing the last (resp. the first) entry from vector $\underline{v}$, and $.^H$ denotes the Hermitian transpose. This leads to the following estimation:

\begin{equation}
\lambda^k(m) = \angle \left(  \underline{\beta}^k(m)_\downarrow^H \underline{\beta}^k(m)_\uparrow \right).
\label{eq:strict_lambda}
\end{equation}

From (\ref{eq:strict_psi}) and (\ref{eq:strict_lambda}) we can derive the full procedure (detailed in Algorithm \ref{al:strict}) of the iterative phase parameters estimation.

\begin{algorithm}[t]
\caption{Strict phase estimation procedure}
\label{al:strict}
\begin{algorithmic}
\STATE \textbf{Inputs}: 
$Y, A, \psi_{ini}, \lambda_{ini}$

\STATE \textbf{Initialization}: \\
$\psi = \psi_{ini}$, $\lambda = \lambda_{ini}$\\
$\hat{Y}^k(f,m) = A^k(f,t_m) e^{ i \psi^k(f)}  e^{ i\lambda^k(m) f}$ \\
$\hat{Y} = \sum_{k=1}^K \hat{Y}^k$\\
$B^k = Y - \hat{Y} + \hat{Y}^k $

\WHILE{stopping criteria not met}

\FOR{$k=1 \text{ to } K \text{, } f=0 \text{ to } F-1 \text{ and } m=1 \text{ to } M-1$}

\item \textbf{Compute $\psi$}

$\psi^k(f) = \angle \left( \sum_{m} B^k(f,m) A^k(f,t_m) e^{- i \lambda^k(m) f} \right)$

\item \textbf{Compute $\underline{\beta}$}

$\beta^k(f,m) = B^k(f,m) e^{- i \psi^k(f)} $\\
$\underline{\beta}^k(m) = [\beta^k(0,m),...,\beta^k(F-1,m)]^T$

\item \textbf{Compute $\lambda$}

$\lambda^k(m) = \angle \left(  \underline{\beta}^k(m)_\downarrow^H \underline{\beta}^k(m)_\uparrow \right)$

\item \textbf{Compute $\hat{Y}$}

$\hat{Y}^k(f,m) = A^k(f,t_m) e^{ i \psi^k(f)}  e^{ i\lambda^k(m) f}$\\
$\hat{Y} = \sum_{k=1}^K \hat{Y}^k$

\item \textbf{Compute $B$}

$B^k = Y - \hat{Y} + \hat{Y}^k $

\ENDFOR
\ENDWHILE

\STATE \textbf{Outputs}:  
$\hat{Y}, \hat{Y}^k, \psi, \lambda$

\end{algorithmic}
\end{algorithm}

\subsection{Relaxed phase constraint}

In this paragraph, we consider a relaxed constraint, which leads to the following cost function:

\begin{multline}
\mathcal{C}_r = \sum_{f,m}| Y(f,m) -  \sum_{k=1}^K  A^k(f,t_m) e^{ i \phi^k(f,t_m)}|^2 \\ + \sigma \sum_{f,m,k} A^k(f,t_m)^2 | e^{ i \phi^k(f,t_m)} - e^{ i \psi^k(f)} e^{i \lambda^k(m) f } |^2,
\label{eq:cost_relax}
\end{multline}
where $\sigma$ is a prior weight which promotes the phase constraint. The minimization of this cost function is performed with a technique similar to the one employed for the strict constraint method. The estimation of the parameter $\lambda^k(m)$ requires the introduction of a new auxiliary variable $\gamma$ which is similar to the variable $\beta$ introduced in the previous paragraph:

\begin{equation}
\gamma^k(f,m) = A^k(f,t_m) e^{ i \phi^k(f,t_m)} e^{- i \psi^k(f)}.
\label{eq:gamma}
\end{equation}

Only one additional step is required for estimating $\phi^k(f,t_m)$, which consists in a calculation identical to (\ref{eq:strict_psi}). The procedure of phase parameters estimation under the relaxed constraint is provided in Algorithm~\ref{al:relax}.


\begin{algorithm}[t]
\caption{Relaxed phase estimation procedure}
\label{al:relax}
\begin{algorithmic}
\STATE \textbf{Inputs}: 
$Y, A, \psi_{ini}, \lambda_{ini}, \phi_{ini}, \sigma$

\STATE \textbf{Initialization}: \\
$\phi = \phi_{ini}$, $\psi = \psi_{ini}$, $\lambda = \lambda_{ini}$\\
$\hat{Y}^k(f,m) = A^k(f,t_m) e^{ i \phi^k(f,t_m)}$ \\
$\hat{Y} = \sum_{k=1}^K \hat{Y}^k$\\
$B^k = Y - \hat{Y} + \hat{Y}^k $

\WHILE{stopping criteria not met}

\FOR{$k=1 \text{ to } K \text{, } f=0 \text{ to } F-1 \text{ and } m=1 \text{ to } M-1$}

\item \textbf{Compute $\phi$}

$\phi^k(f,t_m) = \angle ( B^k(f,m) A^k(f,t_m) $\\
\begin{flushright}
$+ \sigma A^k(f,t_m)^2 e^{ i \psi^k(f)} e^{ i \lambda^k(m) f})$
\end{flushright}

\item \textbf{Compute $\psi$}

$\psi^k(f) = \angle \left(  \sum_m A^k(f,t_m)^2 e^{ i \phi^k(f,t_m)} e^{ - i \lambda^k(m) f} \right)$

\item \textbf{Compute $\underline{\gamma}$}

$\gamma^k(f,m) = A^k(f,t_m) e^{ i \phi^k(f,t_m)} e^{- i \psi^k(f)}$\\
$\underline{\gamma}^k(m) = [\gamma^k(0,m),...,\gamma^k(F-1,m)]^T$

\item \textbf{Compute $\lambda$}

$\lambda^k(m) = \angle \left(  \underline{\gamma}^k(m)_\downarrow^H \underline{\gamma}^k(m)_\uparrow \right)$

\item \textbf{Compute $\hat{Y}$}

$\hat{Y}^k(f,m) = A^k(f,t_m) e^{ i \phi^k(f,t_m)}$\\
$\hat{Y} = \sum_{k=1}^K \hat{Y}^k$

\item \textbf{Compute $B$}

$B^k = Y - \hat{Y} + \hat{Y}^k $

\ENDFOR
\ENDWHILE

\STATE \textbf{Outputs}: 
$\hat{Y}, \hat{Y}^k, \psi, \lambda, \phi$
\end{algorithmic}
\end{algorithm}

\vspace{-1em}

\section{Experimental results}
\label{sec:exp}

In this section we present some experiments conducted to evaluate the potential of our method. For all experiments, signals are sampled at $F_s = 11025$ Hz. The STFT is computed using a $512$ sample-long normalized Hann window with $75 \%$ overlap. The model is estimated by running $100$ iterations of our two algorithms (the performance is not further improved beyond).

The MATLAB Tempogram Toolbox~\cite{Grosche2011} is used to estimate the onset frames. We then extract the onset matrix $Y$ from the full data matrix $X$. The PEASS Toolbox~\cite{Emiya2011} is used to evaluate the source separation quality. It computes the following energy ratios: the SIR (signal to interference ratio) that measures the rejection of interferences, the SAR (signal to artifact ratio) for the rejection of artifacts, and the SDR (signal to distortion ratio) for the global quality.

\subsection{Onset phase estimation}

In this experiment, we seek to estimate the onset phase parameters on mixtures of two sources. For all mixtures, each source is successively observed alone, then the two sources are simultaneously activated. Algorithms are tested on model-built data (i.e. based on (\ref{eq:mod_mix})) and on mixtures of piano notes from the MAPS database~\cite{Emiya2010a}. We test both the strict and the relaxed algorithms, for various values of the parameter $\sigma$. Magnitudes values $A$ are assumed to be known, in order to specifically inquire about the quality of phase estimation. These algorithms are compared to the traditional Wiener filtering approach. For all these methods, we compute the estimation error $\frac{1}{K}\sum_k ||Y^k-\hat{Y}^k||_\mathcal{F}$ where $\hat{Y}^k$ is the estimated $k$-th source within onset frames and $||.||_\mathcal{F}$ denotes the Frobenius norm. Results are averaged over $30$ signals for each dataset and presented in Figure~\ref{fig:attack_phase}.

\begin{figure}
\center
\includegraphics[scale=0.45]{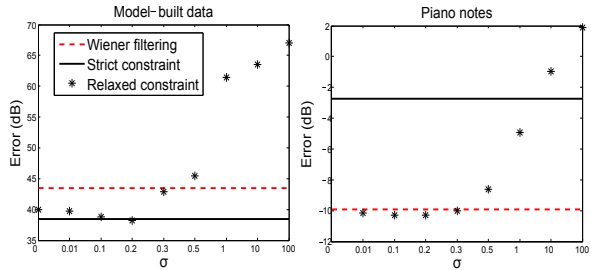}
\caption{Estimation error between data and estimated mixture. Model-built data (left) and piano notes mixture (right).}
\label{fig:attack_phase}
\vspace{-0.8em}
\end{figure}

For the model-built data, the algorithm under strict constraint leads to a better phase estimation than Wiener filtering method. Indeed, this method overcomes the issue of TF overlap, leading to an accurate estimation of the phases of the data. The relaxed algorithm also provides interesting results, however the quality of the estimation is highly dependent on the value of the parameter $\sigma$. On realistic data, the strict algorithm does not perform better than Wiener filtering. The strict phase constraint seems too restrictive to lead to an accurate estimation of the parameters.
However, when the relaxation parameter $\sigma$ is properly chosen (around $0.2$ in this case), our method leads to a slightly better phase estimation than the traditional Wiener filtering technique.

Those results demonstrate the potential of a model exploiting phase repetitions between onset frames. Future research could investigate on the automatic calculation of the optimal value for the relaxation parameter $\sigma$.

\subsection{Application to source separation}
We propose to apply our estimation technique to a source separation task. We consider several datasets:

\begin{itemize}
\item[A:] 30 mixtures of two sources, composed of synthetic damped sinusoids. Sources do not overlap in the TF domain.
\item[B:] 30 mixtures of two sources, composed of synthetic damped sinusoids. Sources overlap in the TF domain.
\item[C:] A $1.57$ second-long MIDI audio excerpt. It is composed of several occurrences of three bass notes, three keyboard notes, and one guitar chord.
\end{itemize}

For mixtures in datasets A and B, each source is successively observed alone, then both sources are activated simultaneously.


Onset phase estimation is performed with the relaxed algorithm and $\sigma = 0.2$. Then, the linear unwrapping algorithm~\cite{Magron2015a} is applied to complete the phase restoration over time frames. This method will be referred to as \textbf{RePU} (\textbf{Re}peating \textbf{P}hase with \textbf{U}nwrapping). It is compared to the traditional Wiener filtering approach.

\begin{table}
\center
\begin{tabular}{c|c||c|c|c}
Dataset & Method & SDR & SIR & SAR \\
\hline
\multirow{2}{*}{A}
& Wiener  & $\mathbf{29.3}$  & $\mathbf{20.8}$ & $\mathbf{58.6}$ \\
& RePU   & $3.2$  & $9.6$ & $26.1$  \\
 \hline
\multirow{2}{*}{B} 
& Wiener  & $\mathbf{10.5}$  & $7.9$ & $20.9$ \\
& RePU   & $3.2$  & $\mathbf{8.8}$ & $\mathbf{25.2}$  \\
 \hline
 \multirow{2}{*}{C} 
& Wiener  & $\mathbf{-2.3}$  & $-20.6$ & $\mathbf{20.8}$ \\
& RePU   & $-3.2$  & $\mathbf{-16.7}$ & $11.6$  \\
 \hline
\end{tabular}
\caption{Average source separation performance (SDR, SIR and SAR in dB) for various algorithms and datasets}
\label{tab:sep_repu}
\vspace{-0.8em}
\end{table}

Results presented in Table~\ref{tab:sep_repu} on dataset A show a clear superiority of Wiener filtering method over our technique. This was expected because there is no overlap in the TF domain in this dataset. Thus, if a source is active in a TF bin, the phase of this source is exactly equal to the phase of the mixture. However, when overlap occurs in the TF domain (dataset B), our method leads to an increase in interference and artifact rejection. This result points out the potential of our technique for separating overlapping sources in the TF domain. Finally, the test conducted on a realistic musical excerpt (dataset C) shows that our method leads to a slight increase in interference rejection compared to Wiener filtering.  

The MATLAB code related to this work and some sound excerpts are provided on the author web page~\cite{Magron}. An informal perceptive evaluation of the source separation quality suggests that the performance measurement employed in these tests may not be able to capture some properties of the separated signals. For instance, the beat phenomenon cannot be suppressed when the phase is retrieved with Wiener filtering, while our technique dramatically attenuates this phenomenon. However, we sometimes observe a loss in transient definition when using our technique.

\section{Conclusion}
\label{sec:conclu}

The model introduced in this paper is a promising tool for estimating phase parameters of components in complex mixtures within onset frames. The phase repetitions are exploited by means of the modeling of an onset phase as the sum of a reference phase and a delay linearly dependent on the frequency. Estimation is performed with an algorithm relying on either a strict or a relaxed constraint. Experimental results show that a fine tuning of the algorithm parameter leads to a more accurate phase estimation than with the traditional Wiener filtering method. In particular, this technique showed its potential for the source separation task when it is combined with a phase unwrapping method.

Future research will focus on refining this repeating model in order to encompass a variety of complex signals, e.g. sources activated with different nuances and intensities. In addition to modeling repetitions from one onset frame to another, we could model the phase correlations between frequency channels within an onset frame. Finally, the magnitudes of the sources are not known in practice, thus they could be estimated jointly with phase parameters within a phase-constrained complex NMF framework~\cite{Kameoka2009}. Indeed, including a phase constraint in a complex NMF model seems to be a promising approach for exploiting both magnitude and phase repetitions in complex mixtures.


\newpage
\bibliographystyle{IEEEtran}
\bibliography{references_waspaa2015}

\end{sloppy}
\end{document}